\documentclass{jetpl}

\newcommand{\R}{{\vec{\rm R}}}
\newcommand{\xv}{{\vec{\rm x}}}
\newcommand{\ka}{{\vec{\rm k}}}
\newcommand{\y}{{\vec{\rm y}}}
\newcommand{\q}{{\vec{\rm q}}}

\newcommand{\vs}{{\vec{\rm v}}}
\newcommand{\n}{{\vec{\rm n}}}

\newcommand{\jv}{\vec{\rm J}}

\newcommand%
{\MVV}%
[1]%
{{\langle\!\langle #1\rangle\!\rangle}}%
\newcommand%
{\MMV}%
[1]%
{{\left\langle\!\!\!\left\langle #1\right\rangle\!\!\!\right\rangle}}%
\def\bldmth#1{%
\mathchoice
{{\hbox{\boldmath$\displaystyle#1$\unboldmath}}}%
{{\hbox{\boldmath$\textstyle#1$\unboldmath}}}%
{{\hbox{\boldmath$\scriptstyle#1$\unboldmath}}}%
{{\hbox{\boldmath$\scriptscriptstyle#1$\unboldmath}}}%
}
\def\vec#1{\bldmth{#1}}
\title{Pre-asymptotic analysis of scattering problem}
\rtitle{Pre-asymptotic analysis of scattering problem}
\sodtitle{Pre-asymptotic analysis of scattering problem}
\author{S.\,E. Korenblit\/\thanks{and LNP JINR RU-141980, Dubna, Russia; 
e-mail: korenb@ic.isu.ru; 
\indent 
$^{2)}$ and Irkutsk National Research Technical University, Lermontov str. 83, Irkutsk, 
Russia.}, 
S.\,V. Lovtsov,  
A.\,V. Sinitskaya$^{2)}$}
\rauthor{S.\,E. Korenblit, S.\,V. Lovtsov, A.\,V. Sinitskaya}
\sodauthor{S.\,E. Korenblit, S.\,V. Lovtsov, A.\,V. Sinitskaya}
\address{Irkutsk State University, Gagarin bl-vd, 20, Irkutsk, 664003, Russia}
\abstract{The pre-asymptotic analysis of the multichannel scattering problem for particles with an arbitrary spin and
short-range interactions has been presented. The complete operator-valued dependence of the scattered differential flux on the distance to the target exactly consistent with the unitarity condition has been obtained.
}
\begin{document}
\UseRawInputEncoding 
\maketitle

{\bf 1. Introduction.} 
As known, because of the local conservation of the
current density, the radial flux of particles emitted into
a given solid angle is independent of the distance $R$ even from an anisotropic point stationary source of
classical particles, light rays, or an incompressible liquid.

In the wave picture, such an independence is valid only for the flux of a spherical divergent 
(convergent) wave $f e^{\pm ikR}/R$ and gives the same inverse-square law  $R^{-2}$ for the 
event rate and the independence of the differential scattering cross section $|f|^2$ from $R$
\cite{Newt,T,LL}. We analyze the possible violation of this law. Being a purely wave effect, 
it occurs because the exact scattered wave is nonspherical, i.e., because of the next,
pre-asymptotic, terms of its asymptotic expansion in $R^{-S}$. Such an expansion is obtained 
in this work in the explicit operator form for all orders $S\geqslant 1$  taking into
account the conservation of the corresponding current.

A growing interest in the pre-asymptotic analysis of the scattering problem is due to 
the results of new experiments and their analysis \cite{anom,NN_shk,k_t3} and due to the 
development of the theory itself \cite{Fer_1,Fer_2,k_Sn,FKS}. In contrast to the long-range 
Coulomb potential, which allows only the nonasymptotic analysis \cite{Fer_1,Fer_2} based on 
the exact Coulomb solutions, an arbitrary short-range local or nonlocal interaction allows the 
pre-asymptotic analysis based on exact free solutions  \cite{k_Sn, FKS}. For example,
such an analysis is possible for the scattering of neutrons on nuclei with the inclusion of 
all neutral channels of such a reaction \cite{Newt,T,LL}. Further, the results of
the pre-asymptotic analysis of single-channel scattering performed in \cite{k_Sn,FKS} are generalized and refined to
this case. We use the system of units where $\hbar=1$. 

{\bf 2. Asymptotic expansion of wave function.} The asymptotic expansion of the wave function 
follows from the operator expansion of the free Green
function. Considering vectors and $\vec{\rm R}=R\n$, 
$\xv=r \vs$ and $\n=(\sin\vartheta\cos\varphi,\sin\vartheta\sin\varphi,\cos\vartheta)$;   
operators $\vec{\nabla}_{\vec{\rm R}}=\n\partial_{R}+R^{-1}\vec{\vec{\heartsuit}}_{\n}$, 
where in spherical basis
$\vec{\vec{\heartsuit}}_{\n}=(0,\partial_\vartheta,(\sin\vartheta)^{-1}\partial_\varphi)$, 
squared ${\cal L}_{\n}\equiv\vec{L}^2_{\n}=-\vec{\vec{\heartsuit}}^2_{\n}$ orbital momentum 
operator  $\vec{L}_{\n}=-i\left(\n\times\vec{\vec{\heartsuit}}_{\n}\right)$ and 
$\Lambda_{\n}=\sqrt{{\cal L}_{\n}+\frac 14}-\frac 12$ we have with $|\xv|=r<R$:  
\begin{align}
&
\frac{e^{\pm ik|\vec{\rm R}-\xv|}}{4\pi|\vec{\rm R}-\xv|}=
\frac{\chi_{\Lambda_{\n}}(\mp ikR)}{4\pi R}e^{\mp ik(\n\cdot\xv)}\sim
\frac{e^{\pm ikR}}{4\pi R} \left\{1+\sum^\infty_{S=1}\frac{ \displaystyle 
\prod\limits^S_{\mu=1}\left[{\cal L}_{\n}-\mu(\mu-1)\right]}
{S!(\mp 2ikR)^S}\right\} e^{\mp ik(\n\cdot\xv)}. 
\label{1} 
\end{align}
The equality in (\ref{1})  is a brief operator representation \cite{k_t3,k_Sn}  of the 
multipole expansion of the Green function  \cite{Newt}  involving the following plane wave 
expansion \cite{T}: 
\begin{align}
e^{\mp ik(\n\cdot\xv)}=  
\frac{4\pi}{kr}\sum^\infty_{l=0}i^{\mp l}\,\psi_{l\,0}(kr)\!
\sum^l_{m=-l}Y^m_l(\n)\overset{*}{Y}{}^m_l(\vs).
\label{3} 
\end{align}
Here and below: $Y^m_l(\n)=\langle\n|l,m\rangle$ is a spherical function; regular  
$\psi_{l\,0}(kr)=(2i)^{-1}\left[i^{-l}\chi_l(-ikr)-i^l\chi_l(ikr)\right]$ and irregular 
solutions of the free radial equation:  
\begin{align}
\chi_l(z)\equiv \left(\frac{2z}{\pi}\right)^{1/2}\!\!
K_{l+{\frac 12}}(z), \quad \; z=\mp ikr,     
\label{4} 
\end{align}
where $K_\lambda(z)$ \cite{gr} is the modified Bessel function of the third kind 
(Macdonald function), are expressed in terms of elementary functions \cite{T,LL,gr}: 
\begin{align}
&
\chi_l(z) \underset{l=\,{\rm int}}{\longmapsto}
e^{-z}\sum^l_{S=0}\frac{(l+S)!}{S!(l-S)!(2z)^S}, \quad z=\mp ikR, 
\label{5} \\
&
\mbox{at: }\;\,\chi_{\Lambda_{\n}}(\mp ikR)Y^m_l(\n)=\chi_l(z)Y^m_l(\n).
\label{5_0}
\end{align}
The series in the parameter $S$ in (\ref{1}) for ${\cal L}_{\n}\mapsto l(l+1)$ is the known asymptotic 
expansion  \cite{gr} of the function  (\ref{4}),  which becomes the finite sum (\ref{5}) 
at integer $l\leftarrowtail\Lambda_{\n}$. 

Following \cite{k_Sn}, we substitute relations (\ref{1}) into Lippmann-Schwinger equation for 
the scattering wave function without rearrangement in the center-of-mass system \cite{Newt}:  
\begin{align}
&
\Psi^{\pm}_{a}(\ka_\alpha \alpha,\R;\vec{\rm a})=
\Psi^0_{a}(\ka_\alpha \alpha,\R;\vec{\rm a})+
\int\!d^3{\rm x}\!\int\!d^3{\rm y}\langle\vec{\rm a}|\widehat{G}^{\pm}_{a}(E;\R,\xv)
\widehat{V}^{a}(\xv,\y)\Psi^{\pm}_a(\ka_\alpha \alpha,\y;),
\label{6}
\end{align}
with the reduced mass ${\rm m}_a$, for the free Green function $\widehat{G}_{a}^{\pm}$ and 
interaction $\widehat{V}^{a}$ operators in the subspace of states of the targets  in  $\alpha(a)$  
channels with energies $\varepsilon_{\alpha(a)}$, and incident  external particle $a$ momentum 
$k_{\alpha(a)}=[2{\rm m}_a(E-\varepsilon_{\alpha})]^{1/2}$ and 
$\ka_\alpha=k_\alpha\vec{\kappa}$:  
\begin{align}
&
\Phi_{\beta(a)}(\vec{\rm a})=\langle\vec{\rm a}|\beta(a)\rangle, 
\label{7} \\
&
\langle\vec{\rm a}|\widehat{G}_{a}^{\pm}(E;\R,\xv)|\vec{\rm a}^\prime\rangle=
-\frac{{\rm m}_a}{2\pi}\sum_{\beta(a)}
\frac{e^{\pm ik_{\beta}|\R-\xv|}}{|\R-\xv|}
\Phi_{\beta(a)}(\vec{\rm a})\Phi^\dagger_{\beta(a)}(\vec{\rm a}^\prime), 
\label{8} 
\end{align}
and similarly for $\widehat{V}^{a}$. The vectors $\R=\R_a,\xv,\y$ specify the coordinates of the 
external particle $a$, and $\vec{\rm a}^{\prime\prime},\vec{\rm a}^\prime,\vec{\rm a}$ 
are sets of vectors of the relative coordinates of particles of the target $(bc)$ in its different states related to the
channels $\alpha(a),\beta(a)$ with the $a$-th external particle corresponding to the separation of the total Hamiltonian
into the free Hamiltonian  $H^{a}$ of the $\alpha(a)$ channels
with the Green function $G_{a}^{\pm}(E)=\left[E\pm i0-H^a\right]^{-1}$ (\ref{8}) and the interaction 
$V^{a}$ in them between the $a$-th particle and target: 
\begin{align}
&  
H=H^{a}+V^{a}, \quad
H^{a}=\vec{\rm P}^2_a/(2{\rm m}_a)+\widehat{H}^{(\vec{\rm a})}_{{bc}},
\label{9} \\
&
H\Psi^{\pm}_{a}(\ka_\alpha \alpha,\R;\vec{\rm a})=
E\Psi^{\pm}_{a}(\ka_\alpha \alpha,\R;\vec{\rm a}), 
\label{9_0} \\
&
H^{a}\Psi^{0}_{a}(\ka_\alpha \alpha,\R;\vec{\rm a})=
E\Psi^{0}_{a}(\ka_\alpha \alpha,\R;\vec{\rm a}), 
\label{10} \\
&
\widehat{H}^{(\vec{\rm a})}_{{bc}}\Phi_{\beta(a)}(\vec{\rm a})=\varepsilon_{\beta}
\Phi_{\beta(a)}(\vec{\rm a}),\;\mbox{ where at}\; E>\varepsilon_{\alpha}: 
\label{10_0} \\
&
\Psi^{0}_{a}(\ka_\alpha \alpha,\R;\vec{\rm a})=(2\pi)^{-3/2}
e^{i(\ka_\alpha\cdot\R)}\Phi_{\alpha(a)}(\vec{\rm a}). 
\label{11}
\end{align}
Integrals over the internal variables $\vec{\rm a}^\prime$ of the target states (\ref{7}), 
entering in (\ref{6}) implicitly \cite{Newt}, are given explicitly in (\ref{13}). 
The separation of the total Hamiltonian in (\ref{9}) 
with the free Hamiltonian $H^{a}$ and the Hamiltonian of the target 
$\widehat{H}^{(\vec{\rm a})}_{{bc}}$ 
distinguishes in (\ref{10_0}) the subgroup of channels $\beta(a)$ with excitations of different bound states of the target $(bc)$, which does not 
contain continuous spectrum states corresponding to its decay \cite{T}. 
Nevertheless, the sum (\ref{8}) over $\beta(a)\mapsto\beta$ \cite{Newt,T,JZ} includes it in this
complete and orthonormalized system of eigenfunctions (\ref{10_0})  and, thereby, includes it in the system of
functions  (\ref{11}). 

The channel indices $\alpha, \beta$ and the variables $\vec{\rm a}$ in (\ref{6})--(\ref{11}) can include discrete indices of spin
degrees of freedom of the system \cite{Newt}. Choosing a common fixed quantization axis for all spins, we can
expand below the scattering amplitude 
$f^\pm_{\beta\alpha}(k_\beta\n;\ka_\alpha)$ only in the spherical functions $Y^m_l(\n)$  
with the quantum numbers and variables of the final states (see \cite{Newt}). 

In (\ref{6}), the matrix elements ${\cal W}^{a(\pm)}_{\beta\alpha}(\xv)$ are given by the
expression:  
\begin{align}
&
{\cal W}^{a(\pm)}_{\beta\alpha}(\xv)=
\!\int\!d^3{\rm y}\!\int\!d^3{\rm a}^\prime\Phi^\dagger_{\beta}(\vec{\rm a}^\prime)
\!\int\!d^3{\rm a}^{\prime\prime}
\langle\vec{\rm a}^\prime|\widehat{V}^{a}(\xv,\y)|\vec{\rm a}^{\prime\prime}\rangle
\Psi^{\pm}_a(\ka_\alpha \alpha,\y;\vec{\rm a}^{\prime\prime})(2\pi)^{3/2},
\label{13} 
\end{align}
the interactions $\widehat{V}^{a}$ from (\ref{9}) being either finite functions $r=|\xv|$, 
vanishing at $r>\rho_0$ or decreasing at $r\to\infty$ faster than  
$\widetilde{\rm C}_{\rm N}r^{-\rm N}$ with any power ${\rm N}$. 
Then note that two presumed additions similar to that for one channel case \cite{k_Sn}: 
\begin{align}
&
\Delta_R f^{\pm}_{\beta\alpha}=
\frac{{\chi}_{\Lambda_{\n}}(\mp ik_\beta R)}{2\pi R}{\rm m}_a
\!\!\!\int\limits_{r>R}\!\!d^3{\rm x}\,e^{\mp ik_\beta(\n\cdot\xv)}
{\cal W}^{a(\pm)}_{\beta\alpha}(\xv), 
\nonumber \\
&
\Delta_R{\cal J}^{\pm}_{\beta\alpha}=
-{\rm m}_a\!\!\int\limits_{r>R}\!\! d^3{\rm x}\,
\frac{e^{\pm ik_\beta|\vec{\rm R}-\xv|}}{2\pi|\vec{\rm R}-\xv|}\,
{\cal W}^{a(\pm)}_{\beta\alpha}(\xv),
\label{12}. 
\end{align}
either vanish at $R>\rho_0$ because of the finiteness of the norm 
$||\Psi||\equiv\underset{\y,\vec{\rm a},\alpha}{\sup}|
\Psi^{\pm}_a(\ka_\alpha \alpha,\y;\vec{\rm a})|$, or also decrease faster than $R^{-\rm N}$
with any power ${\rm N}\gg 1$ due to similar \cite{k_Sn} estimations 
\begin{align}
&
\underset{\beta}{\sup} 
\!\!\int\!\!d^3{\rm y}\!\!\int\!\!d^3{\rm a}^{\prime\prime}
\left|\int\!\!d^3{\rm a}^\prime\Phi^\dagger_{\beta}(\vec{\rm a}^\prime)
\langle\vec{\rm a}^\prime|\widehat{V}^{a}(\xv,\y)|\vec{\rm a}^{\prime\prime}\rangle\right|
\leqslant \frac{{\rm C}_{\rm N}}{r^{\rm N}},  
\qquad \widetilde{\rm C}_{\rm N}={\rm C}_{\rm N}||\Psi||(2\pi)^{3/2}\!,   
\nonumber \\ 
&
|\Delta_R{\cal J}^\pm_{\beta\alpha}|<
\frac{2{\rm m}_a\,\widetilde{\rm C}_{\rm N}}{({\rm N}-2)R^{{\rm N}-2}}, \qquad 
|\Delta_R f^\pm_{\beta\alpha}|<
\frac{2{\rm m}_a\,\widetilde{\rm C}_{\rm N}}{({\rm N}-3)R^{{\rm N}-2}}
\left[1+ O(R^{-1})\right]. 
\label{15}
\end{align}
As a result the indicated substitution gives the explicit operator form of the asymptotic 
expansion of the scattering wave function without rearrangement:
\begin{align}
&
\Psi^{\pm}_{a}(\ka_\alpha \alpha,\R;\vec{\rm a})\sim (2\pi)^{-3/2}\sum_{\beta(a)}
\Phi_{\beta(a)}(\vec{\rm a})
\left\{\delta_{\beta\alpha}e^{i(\ka_\alpha\cdot\R)}+
\frac{{\chi}_{\Lambda_{\n}}(\mp ik_\beta R)}{R}\,
f^\pm_{\beta\alpha}(k_\beta\n;\ka_\alpha)\right\}, 
\label{16} \\
&
f^\pm_{\beta\alpha}(k_\beta\n;\ka_\alpha)=-\,\frac{{\rm m}_a}{2\pi}\! 
\int\!\!d^3{\rm x} e^{\mp ik_\beta(\n\cdot\xv)}
{\cal W}^{a(\pm)}_{\beta\alpha}(\xv), 
\label{17}
\end{align}
which is expressed here in terms of only the physical scattering amplitude (\ref{17}) on the 
energy shell determining the differential and total scattering cross section from the 
$\alpha(a)$ channel to the $\beta(a)$ channel \cite{Newt,T,JZ}: 
\begin{align}
&
\frac{d\sigma_{\beta\alpha}}{d\Omega(n)}=
\frac{k_{\beta}}{k_{\alpha}}
\left|f^\pm_{\beta\alpha}(k_\beta\n;\ka_\alpha)\right|^2, 
\label{18} \\
&
\sigma_{\beta\alpha}=
\frac{k_{\beta}}{k_{\alpha}}\int\!\!d\Omega(\n)
\left|f^\pm_{\beta\alpha}(k_\beta\n;\ka_\alpha)\right|^2. 
\label{19}
\end{align}
The contribution from the incident plane wave (\ref{11}) is absent in (\ref{16}) for all 
inelastic channels with  $\beta\neq\alpha$. The same consideration and estimates for 
collisions with rearrangement lead to expressions similar to (\ref{16}), (\ref{17}), where, 
however, the ``incident'' wave more complex than  (\ref{11}) and the scattered wave in 
(\ref{16}) correspond to different targets and the contribution from the ``incident'' wave 
in the  $\alpha(a)$ channel is 
orthogonal to the bound states of the final target in the $\beta(b)$ channel \cite{Newt,T}. 
Expansion (\ref{1})  is now performed in (\ref{16}) in  powers of the distance 
$R\mapsto {\cal R}={\cal R}_b$ between another final target $(ac)$ and another final  
external particle $(b)$ scattered in a different direction $\vec{\cal R}={\cal R}\vec{\nu}$ in 
the $\beta(b)$ final channel. 
That is,  in (\ref{8}), (\ref{13}), (\ref{17}) at $\alpha=\alpha(a)$, 
$\Psi^{\pm}_{a}(\ka_\alpha \alpha,\R;\vec{\rm a})\equiv 
\widetilde{\Psi}^{\pm}_{a}(\ka_\alpha \alpha,\vec{\cal R};\vec{\rm b})$ we now have 
$\widehat{V}^{a}\mapsto\widehat{V}^{b}$, ${\rm m}_a\mapsto{\rm m}_b$, $\beta\mapsto\beta(b)$, 
$\n\mapsto\vec{\nu}$, and in (\ref{18}), (\ref{19}) the substitution 
$k_\beta/k_\alpha\mapsto \upsilon_\beta/\upsilon_\alpha$ for corresponding velocities 
$\upsilon_\beta=k_\beta/{\rm m}_b$, $\upsilon_\alpha=k_\alpha/{\rm m}_a$ should be made 
\cite{Newt} (cmp. \cite{FKS}).  

In any case, the exact asymptotic expansion of the scattering wave function for a short-range potential is
obtained by the simple replacement $e^{\pm ik_\beta R}\mapsto{\chi}_{\Lambda_{\n}}(\mp ik_\beta R)$ of the exponential in the expression 
for spherical wave by the operator-valued function (\ref{1}) acting only  on the angular 
variables  $\n$ (or $\vec{\nu}$) of the corresponding scattering amplitude (\ref{17}).  

Considering the expression in the curly brackets in (\ref{16}) as the asymptotic expansion of the function 
$\eta^\pm_{\beta\alpha}(\R)$, we write   
\begin{align}
&
\Psi^{\pm}_{a}(\ka_\alpha \alpha,\R;\vec{\rm a})= (2\pi)^{-3/2}\sum_{\beta(a)}
\Phi_{\beta(a)}(\vec{\rm a})\eta^\pm_{\beta\alpha}(\R),
\label{20} \\
&
\eta^\pm_{\beta\alpha}(\R)\sim \delta_{\beta\alpha}e^{i(\ka_\alpha\cdot\R)}+
\frac{e^{\pm ik_\beta R}}{R} \left[f^\pm_{\beta\alpha}(k_\beta\n;\ka_\alpha)+
\sum^\infty_{S=1}
\frac{h^\pm_{S}(k_\beta\n;\ka_\alpha)}{(\mp 2ik_\beta R)^S}\right],
\label{21} \\ 
&
h^\pm_{S}(k_\beta\n;\ka_\alpha)=     
\frac{1}{S!}\prod\limits^S_{\mu=1}\left[{\cal L}_{\n}-\mu(\mu-1)\right]
f^\pm_{\beta\alpha}(k_\beta\n;\ka_\alpha).  
\nonumber  
\end{align}
This means that the coefficients $h^\pm_{S}$ of its asymptotic expansion (\ref{21})  
are observable together with $f^\pm_{\beta\alpha}$. 

{\bf 3. Scattered differential flux, unitary condition, and optical theorem.}
Following \cite{T,k_Sn}, we consider the radial flux element through a 
small element of the spherical surface $\n R^2 d\Omega(\n)$ for the off-diagonal current 
density $\jv_{\gamma\alpha}(\q_\gamma,\ka_\alpha,\R;\vec{\rm a}^\prime,\vec{\rm a})$ 
constructed on the wave functions from (\ref{6}), (\ref{9_0}), taken in the form of  (\ref{20}), (\ref{16}) at 
$\R=R\n$, $\q_\gamma=k_\gamma\vs$, $\ka_\alpha=k_\alpha\vec{\kappa}$,  
$\overset{\leftrightarrow}{\partial}_{\!R}=
\bigl(\n\cdot\overset{\leftrightarrow}{\vec{\nabla}}_{\R}\bigr)=
\overset{\rightarrow}{\partial}_{\!R}-\overset{\leftarrow}{\partial}_{\! R}$. 
Because of the current conservation $\left(\vec{\nabla}_{\R}\cdot
\jv_{\gamma\alpha}(\q_\gamma,\ka_\alpha,\R;\vec{\rm a}^\prime,\vec{\rm a})\right)=0$  
on the equations of motion (\ref{9})--(\ref{10_0}) the total flux through any closed surface 
should be zero,  whereas the radial flux
element is
\begin{align}
&
R^2 d\Omega(\n)\left(\n\cdot
\jv_{\gamma\alpha}(\q_\gamma,\ka_\alpha,\R;\vec{\rm a}^\prime,\vec{\rm a})\right)=
R^2 d\Omega(\n)
\frac 1{2i}\!\left[\overset{*}{\Psi}{}^+_a(\q_\gamma \gamma,\R;\vec{\rm a}^\prime)
\overset{\leftrightarrow}{\partial}_{\! R}
\Psi^{+}_{a}(\ka_\alpha \alpha,\R;\vec{\rm a})\right]\!=
\label{23} \\
&
=\!\frac{R^2 d\Omega(\n)}{(2\pi)^3}\!\sum_{\underline{\beta}}\sum_{\beta}\!
\Phi^\dagger_{\underline{\beta}}(\vec{\rm a}^\prime)\Phi_{\beta}(\vec{\rm a})
\frac 1{2i}\!\left[\overset{*}{\eta}{}^+_{\underline{\beta}\gamma}(\R)
\overset{\leftrightarrow}{\partial}_{\! R}\eta^+_{\beta\alpha}(\R)\right]\!.  
\nonumber   
\end{align}
Here and below, arrows over symbols indicate the
direction of action of operators, and $z_\beta=-ik_\beta R$, etc. 

Substitution (\ref{16}) gives the flux element (\ref{23}) as the sum of three 
terms having a clear physical meaning:
\begin{align}
&
\frac {R^2}{2i}\!\left[\overset{*}{\eta}{}^+_{\underline{\beta}\gamma}(\R)
\overset{\leftrightarrow}{\partial}_{\!R}\eta^+_{\beta\alpha}(\R)\right]\sim
\underbrace{\delta_{\underline{\beta}\gamma}\delta_{\beta\alpha}
\frac{R^2}{2}(\n\cdot(\ka_\alpha+\q_\gamma)) 
e^{iR(\n\cdot(\ka_\alpha-\q_\gamma))}}_{<1>}+
\label{24} \\ 
&
+ \underbrace{\frac{1}{2i} \overset{*}{f}{}^+_{\underline{\beta}\gamma}
(k_{\underline{\beta}}\n;\q_\gamma)
\Bigl[\chi_{\overset{\leftarrow}{\Lambda}{\!}_{\n}}(-z_{\underline{\beta}})
\overset{\leftrightarrow}{\partial}_{\!R} 
\chi_{\overset{\rightarrow}{\Lambda}{\!}_{\n}}(z_\beta)\Bigr]
f^+_{\beta\alpha}(k_\beta\n;\ka_\alpha) }_{<2>} +
\nonumber \\
&
+\underbrace{
\frac {i}{2}\left\{\biggl(\delta_{\underline{\beta}\gamma}e^{z_\gamma(\n\cdot\vs)}
\left[z_\gamma(\n\cdot\vs)+1-z_\beta\frac{\partial}{\partial z_\beta}\right]
\chi_{\overset{\rightarrow}{\Lambda}{\!}_{\n}}(z_\beta)
f^+_{\beta\alpha}(k_\beta\n;k_\alpha\vec{\kappa})\!\biggr)-
\left(\substack{ \displaystyle  \vs \rightleftharpoons \vec{\kappa}, \\ 
\displaystyle \gamma\rightleftharpoons \alpha,}\;\;
\beta\rightleftharpoons \underline{\beta}\right)^{*}\right\}  }_{<3>}\!. 
\nonumber 
\end{align}
As usual \cite{T,LL}, the first term $<\!1\!>$ corresponds to the incident fluxes, the 
second one $<\!2\!>$ now describes the scattered fluxe, and the third one $<\!3\!>$ 
corresponds to their interference. Here and below, we use the following elementary identities for 
derivatives and Wronskians \cite{T} of arbitrary functions $\psi,\phi,w,g$ and the functions 
$\chi_l$ (\ref{4}) of $z$: 
\begin{align}
&
\Bigl[w(z)\psi(z)\overset{\leftrightarrow}{\partial}_{\!z}g(z)\phi(z)\Bigr]=
\psi(z)\phi(z)\Bigl[w(z)\overset{\leftrightarrow}{\partial}_{\!z}g(z)\Bigr]+
w(z)g(z)\Bigl[\psi(z)\overset{\leftrightarrow}{\partial}_{\!z}\phi(z)\Bigr], \qquad 
\Bigl[\phi(z)\overset{\leftrightarrow}{\partial}_{\!z}\phi(z)\Bigr]=0,
\label{25_0} \\  
&
z\Bigl[w(z)\left(\overset{\leftrightarrow}{\partial}_{\!z}+\frac{1}{z}\right)
\frac{g(z)}{z}\Bigr]=\Bigl[w(z)\overset{\leftrightarrow}{\partial}_{\!z}g(z)\Bigr], 
\label{25} \\ 
&
zc e^{zc}=z\partial_z e^{zc}, \qquad 
\Bigl[\chi_l(z)\overset{\leftrightarrow}{\partial}_{\!z}\chi_l(-z)\Bigr]=2. 
\label{26}  
\end{align}
Taking into account the completeness and orthonormality of the system of target functions given by Eqs. (\ref{7}), 
(\ref{10_0}), 
\begin{align}
&
\int\!d^3{\rm a}\Phi^\dagger_{\underline{\beta}}(\vec{\rm a})\Phi_{\beta}(\vec{\rm a})=
\delta_{\underline{\beta}\beta}, 
\label{27}  
\end{align}
these identities determine the result of the calculation of the total flux from (\ref{23}) and (\ref{24}) integrating over 
these functions at $\vec{\rm a}^\prime=\vec{\rm a}$.  
The multiplication by $\delta_{\underline{\beta}\beta}$ and summation over all channels in
(\ref{23}) reduce the contribution from the integral $\int\!d\Omega(\n)\!<\!1\!>$ at 
$\gamma=\alpha$ to integral (42) in \cite{k_Sn}, which is zero. Since the operator $\Lambda_{\n}$ is self-adjoint on the unit sphere and because
of identity (\ref{26}), the same operations reduce the contribution to the result of integration (\ref{23}) 
from the term $<\!2\!>$  to the form of the right-hand side of the unitarity
condition 
\cite{Newt,T,LL}:  
\begin{align}
&
\longmapsto \!\sum_{\beta(a)}k_\beta \!\int\!\! d\Omega(\n)
\overset{*}{f}{}^+_{\beta\gamma}(k_{\beta}\n;\q_\gamma)
f^+_{\beta\alpha}(k_\beta\n;\ka_\alpha).  
\label{28}  
\end{align}
Here, as above, the asterisk $*$ stands for complex conjugation and/or Hermitian conjugation in the case of spin 
indices.Therefore, the same operations should transform the contribution $<\!3\!>$ to the left-hand side of the 
unitary condition. Indeed,  the
integral $\int\!d\Omega(\n)\!<\!3\!>$
at  $\underline{\beta}=\beta=\gamma$ (or $\alpha$) 
 can be determined using the following always possible expansion
of the scattering amplitude mentioned in Section 2:
\begin{align}
&
f^+_{\gamma\alpha}(k_\gamma\n;k_\alpha\vec{\kappa})=\sum^\infty_{l=0}\sum^l_{m=-l}
Y^m_l(\n)B^{lm}_{\gamma\alpha}(k_\gamma;k_\alpha;\vec{\kappa}), 
\label{29}  
\end{align}
which effectively changes $f^+_{\gamma\alpha}(\n)\mapsto Y^m_l(\n)$ in the last line of 
(\ref{24}). Taking into account (\ref{5_0}), this
expansion provides the partial contribution to
$\int\!d\Omega(\n)\!<\!3\!>$ in the form (omitting $B^{lm}_{\gamma\alpha}$): 
\begin{align}
&
\mapsto \frac i2\left\{\!\left(\!z_\gamma\!\left[\chi_l(z_\gamma)\!
\left(\overset{\leftrightarrow}{\partial}_{\!z_\gamma}\!+\!\frac 1{z_\gamma}\right)
\frac{A(z_\gamma)}{z_\gamma}\right]\!\right)- 
\left(\substack{ \displaystyle  \vs \rightleftharpoons \vec{\kappa} \\ 
\displaystyle \gamma\rightleftharpoons \alpha}\right)^{*}\!\right\}, 
\nonumber \\
&
\mbox{where }\;\, \frac{A(z_\gamma)}{z_\gamma}=
\!\int\!\!d\Omega(\n)e^{z_\gamma(\n\cdot\vs)}Y^m_l(\n).
\label{30}  
\end{align}
According to expansion (\ref{3}) and the orthonormality of the spherical functions,  
\begin{align}
&
\frac{A(z_\gamma)}{z_\gamma}=2\pi Y^m_l(\vs) 
\left[\frac{\chi_l(-z_\gamma)-(-1)^l\chi_l(z_\gamma)}{z_\gamma}\right]. 
\label{31}  
\end{align}
A different method of calculation of the contribution
from the term $<\!3\!>$  in the spinless case 
\cite{k_Sn} shows that $\chi_l(-z_\gamma)$  and $\chi_l(z_\gamma)$ describe  interference 
in the forward and backward directions with $(\n\cdot\vs)=1$ and $(\n\cdot\vs)=-1$, respectively. 
Then, according to (\ref{25}) and 
(\ref{25_0}), (\ref{26}), interference in the backward direction
is absent in all orders of $R^{-S}$ and the contribution from interference in the forward direction  in all orders of $R^{-S}$ 
 is again reduced to a spherical function, which
thus results in the amplitude given by
(\ref{29})  depending already on $\vs$ and, finally, in the left-hand side of
the unitarity 
condition \cite{Newt,T} (with the minus sign). That is, according to (\ref{31}), 
\begin{align}
&
\int\!d\Omega(\n)\!<\!3\!>\,\mapsto\,-\,\frac{4\pi}{2i} Y^m_l(\vs)B^{lm}_{\gamma\alpha}
\,\longmapsto 
-\,\frac{4\pi}{2i}\left[f^+_{\gamma\alpha}(\q_\gamma;\ka_\alpha)-
\overset{*}{f}{}^+_{\alpha\gamma}(\ka_\alpha;\q_\gamma) \right]. 
\label{32}  
\end{align}
The vanishing of the total flux as the sum of (\ref{28}) and (\ref{32}) gives the unitarity
condition \cite{Newt,T, LL} and,  
taking into account (\ref{18}), (\ref{19}), leads at $\gamma=\alpha$, $\vs=\vec{\kappa}$  to the optycal theorem. 
Therefore, the unitarity condition and optical theorem are applicable not only
in the far-field region but also in the near-field one  at 
finite distances from the scattering target. The 
{\sl differential flux} {\sl scattered} without  rearrangement from the $\alpha(a)$  channel, 
which explicitly depends on this distance, is determined by the term $<\!2\!>$ in  (\ref{24}) as
\begin{align}
&
\frac{d\Sigma_\alpha(R)}{d\Omega(\n)}=\sum_{\beta(a)}\frac{1}{2ik_\alpha}
\overset{*}{f}{}^+_{\beta\alpha}(k_{\beta}\n;\ka_\alpha)
\Bigl[\chi_{\overset{\leftarrow}{\Lambda}{\!}_{\n}}(-z_{\beta})
\overset{\leftrightarrow}{\partial}_{\! R} 
\chi_{\overset{\rightarrow}{\Lambda}{\!}_{\n}}(z_\beta)\Bigr]
f^+_{\beta\alpha}(k_\beta\n;\ka_\alpha).
\label{33} 
\end{align}
This quantity is experimentally measured at finite distances $R$ 
instead of the sum of differential cross sections (\ref{18}), and, at sufficiently large 
values $k_\beta R>1$, it really has the form 
\begin{align}
&
\frac{d\Sigma_\alpha(R)}{d\Omega(\n)}=\sum_{\beta(a)}\frac{k_\beta}{k_\alpha}
\left\{\Bigl|f_{\beta\alpha}(k_\beta\n;\ka_\alpha)\Bigr|^2-  
\frac 1{(k_\beta R)}\,Im\left[f^*_{\beta\alpha}
{\cal L}{}_\n f_{\beta\alpha}\right]+ 
\frac 1{(2k_\beta R)^2}\left[\Bigl|{\cal L}{}_\n f_{\beta\alpha}\Bigr|^2 -
Re\left(f^*_{\beta\alpha}{\cal L}{}^2_\n f_{\beta\alpha}\right)\right]+
\right. \label{34} \\ 
&
+\frac{1}{3(2k_\beta R)^3}Im\Bigl[f^*_{\beta\alpha}{\cal L}{}^3_{\n}f_{\beta\alpha}-
3\left({\cal L}{}_{\n}f_{\beta\alpha}\right)^*{\cal L}{}^2_{\n}f_{\beta\alpha}-  
2f^*_{\beta\alpha}{\cal L}{}^2_{\n}f_{\beta\alpha}\Bigr]+
\nonumber \\
&
+\frac{1}{12(2k_\beta R)^4}\biggl(3\bigl|{\cal L}{}^2_\n f_{\beta\alpha}\bigr|^2+
Re\left[f^*_{\beta\alpha}{\cal L}{}^4_{\n}f_{\beta\alpha}-
4\left({\cal L}{}_{\n}f_{\beta\alpha}\right)^*{\cal L}{}^3_{\n}f_{\beta\alpha}\right]+ 
12\left[Re\left(f^*_{\beta\alpha}{\cal L}{}^2_\n f_{\beta\alpha}\right)-
\bigl|{\cal L}{}_\n f_{\beta\alpha}\bigr|^2\right]-
\nonumber \\
&
\left. 
-8 Re\!\left[f^*_{\beta\alpha}{\cal L}{}^3_{\n}f_{\beta\alpha}\!-
\left({\cal L}{}_{\n}f_{\beta\alpha}\right)^*\!\!{\cal L}{}^2_{\n}f_{\beta\alpha}\right]
\!\biggr)\!+O\!\left((k_\beta R)^{-5}\right)\!\right\}\!\!, 
\nonumber  \\
&
\mbox{where }\;\,f_{\beta\alpha}=f^+_{\beta\alpha}(k_\beta\n;\ka_\alpha), \quad 
\left({\cal L}{}_{\n}f_{\beta\alpha}\right)^*=
f^*_{\beta\alpha}\overset{\leftarrow}{\cal L}{}_{\n},
\nonumber \\
&
\mbox{and only }\;\,\lim_{R\to\infty}\frac{d\Sigma_\alpha(R)}{d\Omega(\n)}=\sum_{\beta(a)}
\frac{d\sigma_{\beta\alpha}}{d\Omega(n)}.
\label{35} 
\end{align}
Unlike the usually used far-field region $k_\beta R\gg 1$, the corrections to the first term in
(\ref{34})  become noticeable long before reaching the near-field region, where $k_\beta R\ll 1$, 
 e.g., near the threshold of each channel. At the same time, all these corrections vanish
automatically in each order in $R^{-S}$ after integration over  $d\Omega(\n)$, 
and, according to (\ref{28}) and (\ref{19}),
\begin{align}
&
\Sigma_\alpha(R)\equiv\int\!d\Omega(\n)\frac{d\Sigma_\alpha(R)}{d\Omega(\n)}=
\sum_{\beta(a)}\sigma_{\beta\alpha}.
\label{36}
\end{align}
According to \cite{k_Sn}, since expansion (\ref{21}) has only asymptotic sense, Eq. (\ref{34}) 
is applicable even to interactions decreasing in (\ref{15}) as $r^{-{\rm N}}$ with 
${\rm N}\geqslant 7$. 
According to \cite{FKS}, the Eq. (\ref{34}) with 
$k_\beta/k_\alpha\mapsto \upsilon_\beta/\upsilon_\alpha$ remains valid, including the first 
relativistic correction $\propto \left(\vec{\rm P}^2_a\right)^2$  to the kinetic  energy 
operator in (\ref{9}). 

For  the same reasons \cite{T}, discussed in Section 2, only the contribution of 
differential fluxes $<\!2\!>$  
of the form of (\ref{33}) holds in (\ref{24}) applied to collisions
with rearrangement, where $k_\beta/k_\alpha\mapsto \upsilon_\beta/\upsilon_\alpha$ in(\ref{34}), 
while the scattering amplitude 
$f^+_{\beta(b)\alpha(a)}(k_{\beta(b)}\vec{\nu};\ka_{\alpha(a)})$ is defined as the coefficient of the
divergent spherical wave in the wave function 
$\Psi(\R)=\widetilde{\Psi}(\vec{\cal R})$ from (\ref{9_0}) and (\ref{6}) at ${\cal R}\to\infty$ 
\cite{Newt,T,JZ}. 
We recall that the determination of the inelastic scattering cross section requires the 
rearrangement of the asymptotic expression for $\Psi(\R)$ even in the single-channel case 
\cite{LL}. 

The substitution of (\ref{29}) into (\ref{33}) gives the corresponding partial expansion for the 
scattered differential flux at $z_\beta=-ik_\beta R$: 
\begin{align}
&
\frac{d\Sigma_\alpha(R)}{d\Omega(\n)}=\!\sum_{\beta(a)}\frac{k_\beta}{k_\alpha}
\sum^\infty_{l,j=0}\sum^l_{m=-l}\sum^j_{\mu=-j}\!
\overset{*}{Y}{}^m_l(\n)Y^\mu_j(\n)\times 
\label{37}  \\
&
\times\!
\overset{*}{B}{}^{lm}_{\beta\alpha}(k_\beta;k_\alpha;\vec{\kappa})
B^{j\mu}_{\beta\alpha}(k_\beta;k_\alpha;\vec{\kappa})\frac 12\!
\Bigl[\chi_{j}(z_{\beta})\overset{\leftrightarrow}{\partial}_{\!z_\beta} 
\chi_{l}(-z_\beta)\Bigr]\!,  
\nonumber
\end{align}
which is certainly consistent with (\ref{36}) taking into account (\ref{26}) and the orthorormality of the spherical functions
$Y^m_l(\n)$. The corresponding analog of (\ref{34}) is obtained by the substitution of the expansion for 
the Wronskian similar to \cite{k_Sn}: 
\begin{align}
&
\frac 12\!\Bigl[\chi_{j}(z_{\beta})\overset{\leftrightarrow}{\partial}_{\!z_\beta} 
\chi_{l}(-z_\beta)\Bigr]\!=\!1+\!\frac{\Delta_{jl}}2\!\!
\int\limits^{\infty}_{z_{\beta}}\!\!\frac{d\zeta}{\zeta^2}
\chi_{l}(-\zeta)\chi_{j}(\zeta)=
1+\Delta_{jl}\sum^{l+j}_{n=0}\frac{A_n(l,j)}{(n+1)(2z_{\beta})^{n+1}},
\label{39}
\end{align}
taken with the necessary accuracy in the powers of $R^{-S}$ at 
\begin{align}
&
\Delta_{jl}=j(j+1)-l(l+1),\quad \Upsilon_{jl}=j(j+1)+l(l+1)\!, 
\nonumber \\
&
\frac 12\!\Bigl[\chi_{j}(z_{\beta})\overset{\leftrightarrow}{\partial}_{\!z_\beta} 
\chi_{l}(-z_\beta)\Bigr]=1+\frac{\Delta_{jl}}{2z_{\beta}}+
\frac{\Delta^2_{jl}}{2(2z_{\beta})^2}+
\label{40}  \\
&
+\frac{\Delta_{jl}}{3(2z_{\beta})^3}\left[\frac{\Delta^2_{jl}}{2}-\Upsilon_{jl}\right]+
\frac{\Delta^2_{jl}}{3(2z_{\beta})^4}
\left[\frac{\Delta^2_{jl}}{8}-\Upsilon_{jl}+\frac 32\right].
\nonumber
\end{align}
The calculation of the coefficients $A_n(l,j)$ shows their
very strong dependence on shifting any of integer parameters $n,l,j>0$ by 1.
 \begin{figure}[htb]
 \centering 
\includegraphics[width=0.47\textwidth]{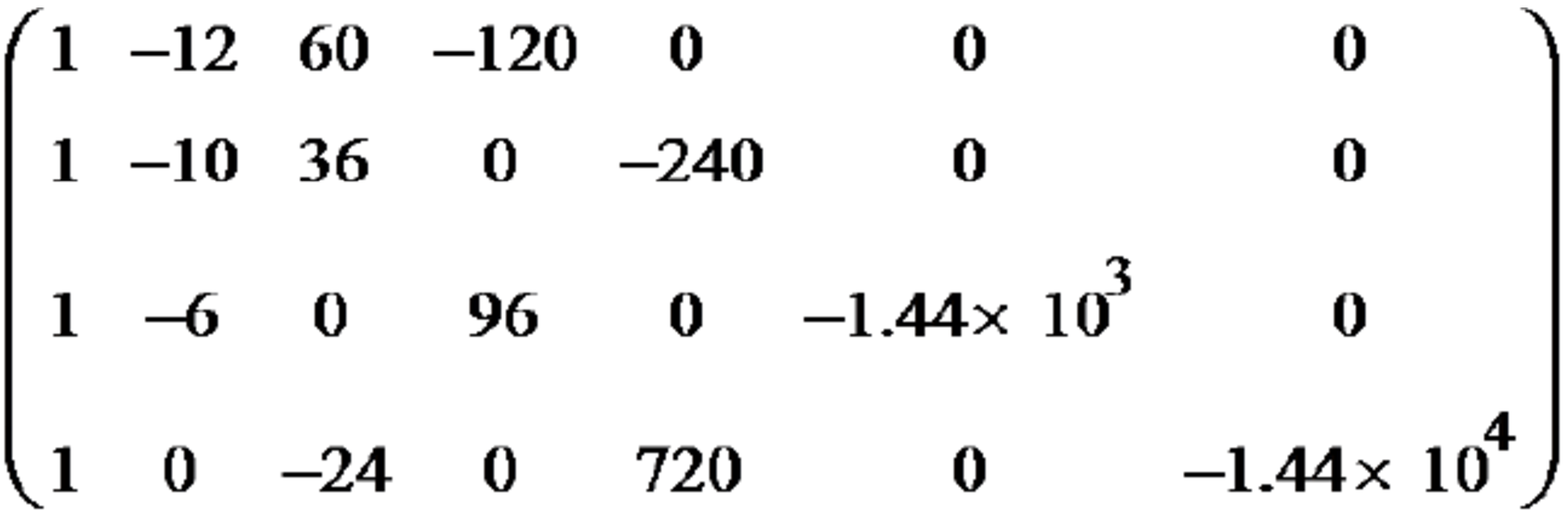}
\vskip 1.5mm 
 \caption{Values of $A_n(l,j)$ (\ref{39}), (\ref{40}) at $l=3$ with $0\leqslant j\leqslant 3$ 
 vertically and $0\leqslant n\leqslant l+j$ horizontally. $A_0(l,j)=1$.}
 \label{fig:1}
 \end{figure}

{\bf Conclusion.} To summarize, using the operator expansion (\ref{1}) of the free Green 
function, we have derived the exact asymptotic expansion given by (\ref{20}), (\ref{21}) for 
wave functions of multichannel scattering of particles with an arbitrary spin for a wide class 
of short-range interactions. The coefficients of this expansion are determined only by the 
physical scattering amplitude  (\ref{17}). 
The derived expansion has confirmed the applicability of the unitarity condition
given by (\ref{28}), (\ref{32}) and of the optical theorem not only in the far-field region 
but also in the near-field one, where the scattered differential flux given by (\ref{33}), 
(\ref{37}), which explicitly depends on the distance $R$, assumes at finite distances $R$ the 
role of the differential scattering cross section given by (\ref{18}), (\ref{35}). 
At quite large values $k_\beta R>1$, this flux can be represented by the first terms of the 
expansions given by Eq. (\ref{34}) or by Eqs. (\ref{37}), (\ref{40}). 

This circumstance makes it possible to expect that the application of these relations 
to processing of the results of scattering experiments with a sufficiently short and variable 
base $R$ will not only give additional information on the corresponding interactions 
\cite{k_t3}, but will also allow estimating finer quantum effects \cite{FKS}. 

We are grateful to V.A. Naumov, D.V. Naumov, M.V. Polyakov, N.V. Ilyin, and A.E. Rastegin 
for stimulating discussions.

\vfill\eject


\end{document}